\documentclass[a4paper,pre, twocolumn,nofootinbib]{revtex4-1}

\usepackage[utf8x]{inputenc}
\usepackage[T1]{fontenc}
\usepackage{lmodern}
\usepackage{siunitx}
\usepackage{xcolor}
\usepackage{xparse}
\usepackage{float}
\usepackage{graphicx}
\usepackage{amsmath}
\usepackage{booktabs}

\usepackage{color}

\begin{document}
 
\title{Extraction du solvant d'un hydrogel par des gouttes de bactéries \textit{B. subtilis}}

\author{Marc Hennes}
\author{Julien Tailleur}
\author{Gaëlle Charron}
\author{Adrian Daerr}
\affiliation{Laboratoire Matière et Systèmes Complexes (MSC) UMR CNRS
  7057, University Paris Diderot, 75205 Paris cedex 13, France}


 



\begin{abstract}

  Nous observons que de petites gouttes de suspension de la bactérie
  \textit{B. subtilis}, déposées sur un gel d'agar, décuplent leur
  volume en une dizaine d'heures, alors que le volume de gouttes sans
  bactérie reste inchangé. L'augmentation de la biomasse due à la
  croissance des bactéries est trop faible pour expliquer le
  gonflement constaté des gouttes, comme le prouve une mesure directe
  de la concentration bactérienne au cours de l'expérience et à
  plusieurs hauteurs au sein de la goutte. Nous montrons que le
  gonflement est provoqué par la présence d'un surfactant produit par
  les bactéries, la \textit{surfactine}, lequel induit un pompage
  d'eau de l'environnement vers la goutte par un effet osmotique
  capillaire. La concentration requise est très faible ($<1$\,mM),
  environ quatre ordres de grandeur plus faible par exemple que la
  concentration de glucose produisant le même gonflement. Cette
  capacité de \textit{B. subtilis} à extraire de l'eau de
  l'environnement contribue probablement à certains modes de migration
  collective comme le \textit{mass swarming}. Elle donne également
  lieu à un mode de déplacement original, indépendant de la motilité
  cellulaire: en associant les effets osmotiques et de mouillage du
  surfactant, \textit{B. subtilis} parvient à faire activement glisser
  la colonie sur des substrats inclinés d'à peine quelques dixièmes de degrés.

\end{abstract}



\maketitle
\noindent
\textbf{Introduction}
\\
Les micro-organismes ont développé d'innombrables mécanismes pour
faciliter leur prolifération \cite{stoodley2004} et la colonisation de
nouveaux environnements \cite{jacob1994,fujikawa1989,harshey2003}. La
raison est sans doute à chercher dans la complexité et la variabilité
spatiale et temporelle de leur environnement. Que ce soit dans un
intestin ou dans le sol, les bactéries se déplacent dans un mélange de
phases solide, liquide et de gaz et leur motilité dépendra
grandement de l'interaction avec les nombreuses surfaces et
d'interfaces présentes. À côté de cette complexité, les études en
laboratoire de la motilité de bactéries se limitent à peu
d'environnements simples et artificiels. En dehors de la nage en
milieu liquide \cite{tuval2005}, on ne trouve presque que des études à
la surface d'hydrogels. Il est donc important de comprendre les
mécanismes physiques en jeu pour pouvoir se prononcer sur l'importance
d'une expérience de laboratoire dans un contexte plus complexe.\\
\indent
Ici, nous décrivons pour la premiere fois un mécanisme
physico-chimique par lequel la bactérie \textit{Bacillus subtilis},
piegée dans une petite goutte deposée sur la surface de l'hydrogel
Agar, est capable d'extraire une quantité importante de solvant,
multipliant ainsi le volume de la goutte jusqu'à atteindre 15 fois le
volume initial. Ce gonflement est proportionnel à la concentration
initiale de bactéries au sein de la goutte, mais est absolument
indépendant de leur capacité à s'auto-propulser. Bien qu'on puisse
s'interroger sur la contribution de la division des bactéries à une
certaine augmentation du volume, nous montrons que dans les conditions
expérimentales présentes, cet effet est négligeable et le gonflement
s'explique par la production d'un tensio-actif sécrété par les
bactéries, la \textit{surfactine}. Pour des gouttes de bactéries
mutantes incapables de produire ce surfactant, aucun gain n'est
visible. Le taux de gonflement dépend fortement des conditions
expérimentales et de la préparation du gel, où l'humidité à
disposition dans l'entourage agit comme facteur limitant. Grâce à des mesures de chromatographie liquide à haute performance, nous montrons que la concentration molaire de surfactine dans les
gouttes est cent à mille fois plus faible que la concentration
d'autres espèces moléculaires, ce qui soulève la question de ce qui
rend la surfactine si efficace pour le pompage osmotique. Nous
terminons par une discussion de l'importance de ce mécanisme pour les
modes migratoires de \textit{B. subtilis}.\\
\\
\noindent
\textbf{R\'esultats} \\
\textit{Description du gonflement de goutte}\quad Une goutte de 2 $\mu$L
provenant d'une culture liquide de \textit{B. subtilis} (densité
optique OD $\approx 0.27$, souche OMG 930  \cite{hamze2011}) est
d\'eposée à la surface d'un gel d'Agar à $0.7\,\%$ contenant des
nutriments (milieu synthétique B), séché bri\`evement (2 min) et incubé
dans une chambre climatique à température et humidité fixe
($\text{T}=30^{\circ}$ C, $\text{RH}=70\, \%$). Pendant la première
heure, un léger étalement radial est visible  \cite{lee2008,banaha2009},
typique pour une solution aqueuse de surfactant --- comportement attendu
car notre souche \textit{B. subtilis} produit de la
\textit{surfactine}, un lipopeptide cyclique
amphiphile  \cite{arima1968,peypoux1999}. Le volume de la
goutte\footnote{Le volume est mesuré grâce à une m\'ethode de
profilométrie de projection de Moiré  \cite{banaha2009}.} augmente dès
le moment du dépôt, de manière monotone et atteint une valeur maximale
de sept à huit fois le volume initial après $\approx 350$ min (Fig. 1\,(a)). En variant la concentration initiale de bactéries (OD pour
\textit{optical density}) jusqu'à la concentration de saturation de
croissance des bactéries $\text{OD} \approx 1.5$, le gonflement peut
atteindre une valeur de $15\,V_i$. Si les bactéries sont enlevées de la
solution par centrifugation et filtrage, et que seul le surnageant SN
est déposé sur le gel, on mesure une augmentation plus faible du
volume de la goutte $V_{\text{max}}^{\rm SN} \approx 3 V_i$. 

\begin{figure*}[ht!]
\centering
\includegraphics[width=.9\textwidth]{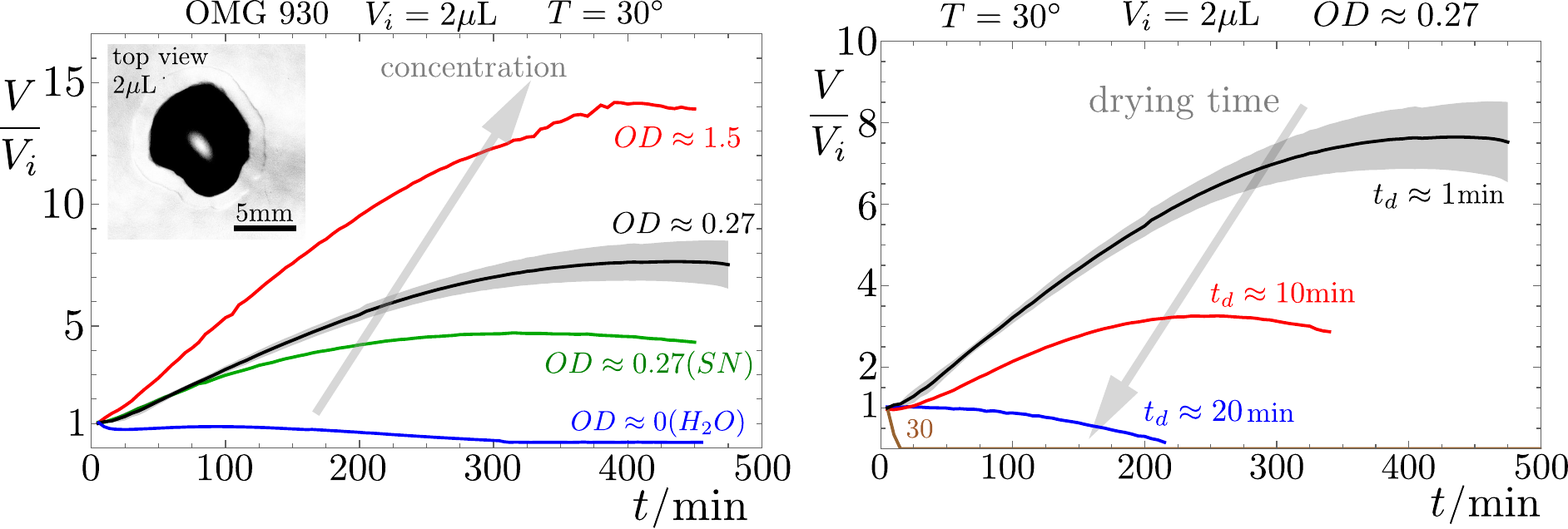}
\caption{(a) Le volume total de gonflement est une fonction de la concentration initiale (OD) de bactéries dans la goutte. Aucun gonflement n'est observable pour des gouttes d'eau pure ou de milieu nutritif B (OD = 0). (b) Le gonflement diminue et disparaît entièrement si le gel est séché plus longtemps à l'air libre avant l'expérience.}
\end{figure*}

Quand le temps de séchage du gel est augmenté (de 1 min à 30 min), l'effet de
pompage est affaibli et la quantité de solvant extrait diminue (Fig. 1 \,(b)). Sur un gel séché pendant $t_{\text{sec}} = 30 \text{min}$,
une goutte de bactérie ($\text{OD} \approx 0.27$) ne gonfle plus du
tout. Ceci s'explique par le fait qu'un tel gel a perdu environ
$100\,V_i$ d'eau par évaporation et n'est donc plus à l'équilibre: le
réseau de polymères devrait être en compression et la pression dans la
phase liquide négative. Pour des gouttes d'eau ou de milieu nutritif
B, une augmentation du volume n'est jamais observée, et un taux
d'évaporation faible dans les coupelles, plus petit que le taux
d'\'evaporation de séchage, fait lentement disparaître ces
gouttes.

\begin{figure*}[ht!]
\centering
\includegraphics[width=0.9\textwidth]{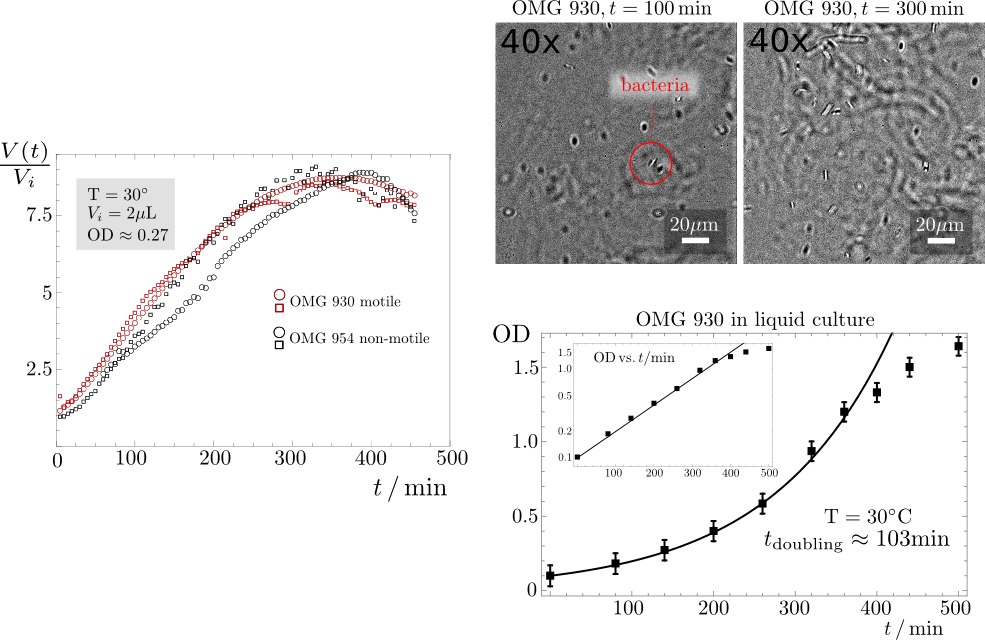}
\caption{(a) La motilité bactérielle, \'evaluée entre la souche OMG 930 motile et la souche OMG 954 non motile, n'influence pas l'accroissement du volume durant les 500 minutes de l'exp\'erience. (b) L'enregistrement des gouttes bactérielles au microscope optique à fort grossissement montre que le pourcentage surfacique des bactéries est tr\`es faible. L'augmentation de la biomasse (avec un temps de doublage de 103 min à $30^{\circ}$ C dans un milieu liquide B, en bas) n'explique donc pas le gonflement des gouttes.}
\end{figure*}

\par \noindent \textit{L'autopropulsion et la biomasse n'influencent pas le
  gonflement}\quad Le gonflement étant absent pour des gouttes d'eau
ou de nutriments, la présence des bactéries est nécessaire pour une
augmentation continue du volume de la goutte. Pour v\'erifier une
\'eventuelle contribution de la motilité individuelle, cruciale pour
le mode de migration de
\textit{swarming}  \cite{fujikawa1989,hamze2011,hamze2009}, nous
comparons les courbes de volume pour des souches motiles (OMG 930) et
non motiles (OMG 954  \cite{hamze2009}) (cf. Fig. 2\,(a)). La deuxième
souche ne se distingue génétiquement de la première que par l'absence
d'un gène qui la rend incapable d'exprimer la protéine
\textit{flagelline} nécessaire au développement des flagelles. Pour
différentes concentrations initiales ($\mathrm{OD}=0.27 \text{ et } \mathrm{OD}=1.5$) nous
observons que les courbes de volume, à nouveau mesurées sur 500 min,
ne diffèrent pas significativement d'une souche à l'autre, montrant
ainsi que la motilité ne joue aucun rôle dans le mécanisme
d'accroissement de volume.

\par Une explication serait alors que l'amassement de bactéries par division cellulaire est responsable de l'augmentation en volume. 
En conditions idéales ($T \approx 40^{\circ}$ C, milieu LB riche en nutriments), le temps de division peut être de 30 minutes  \cite{real2005} ou moins, obtenant ainsi 14 à 15 cycles de r\'eplication pendant le temps de gonflement ($t \approx 400 \text{min}$) d'une goutte. Le nombre absolu de bactéries augmenterait d'un facteur $2^{15}$, correspondant à un volume final de $3.5\, V_i$  \cite{loo2014}. Pour d\'eterminer la contribution biomassique au gonflement, nous enregistrons la goutte à fort grossissement (40$\times$, Fig. 2\,(b), haut) pendant 400 min à différentes hauteurs et calculons le pourcentage surfacique et volumique à travers la goutte. Nous trouvons que ce pourcentage volumique après 7 heures ne dépasse jamais 2\,\% au sein de la goutte et 8\,\% proche de la ligne de contact de bord\footnote{Un pourcentage plus élevé proche de la ligne triple/ligne de contact de la goutte est probablement dû aux flux convectionnels d'\'evaporation dans la goutte, entra\^inant les bactéries vers son bord.}. De plus, nous mesurons directement la concentration bactérienne en solution liquide dans le milieu nutritif minimal B à $30^{\circ}$ C. Ici, nous déterminons un temps de doublage de 90 à 120 minutes (Fig. 2\,(b)), donnant ainsi 4 cycles de croissance en 400 min et un volume final de bactéries de $N\times V_{\text{Bact}} \ll V_i$ largement inférieur au volume initial de la goutte.
\par
\noindent\textit{D\'etermination et quantification de la surfactine comme agent principal}\quad Le comportement de la souche OMG 168 de \textit{B. subtilis}, qui ne produit pas de surfactine  \cite{lesuisse1993}, permet d'identifier ce surfactant comme premier responsable du mécanisme de pompage: dans le cas de gouttes de OMG 168, les courbes de volume ne montrent aucun gonflement et se rapprochent des courbes d'eau pure ou de solution de nutriment B (cf. Fig. 3\,(a)). Nous nous attendons donc à ce que des gouttes de surfactine pure (surfactine de \textit{B. subtilis}, 98\,\% pure, Sigma-Aldrich) affichent une croissance de volume similaire aux gouttes de bactéries produisant de la surfactine. Effectivement, nous observons pour de telles
gouttes un gonflement de magnitude proche des gouttes bactériennes
pour des concentrations entre 0.1 g/L et 0.5 g/L. Ici, une goutte de
surfactine se comporte comme une goutte de surnageant: dans les
premières 100 à 150 minutes, le gain en volume est proche de celui d'une goutte
de bactéries, puis s'att\'enue lentement pour atteindre une valeur 3 à 4
fois au-dessus du volume initial. Afin de vérifier si la quantité de surfactine produite par les bactéries est dans la plage de concentrations dans laquelle les gouttes de surfactines gonflent, nous l'avons quantifié par HPLC (\textit{high precision liquid profilometry}). Cette méthode
spectrographique permet d'identifier et de mesurer les pics d'absorbtion caractéristiques d'une molécule pour remonter à sa concentration (dans un premier temps en unités arbitraires AU). Un étalonnage par des gouttes de concentration connue est nécessaire pour calculer une concentration absolue. La figure 3\,(b) montre que la concentration de surfactine dans la
goutte bactérienne est initialement proche de 0.1 g/L pour
une $\text{OD} \approx 0.27$ et de 0.25 g/L pour une
$\text{OD} \approx 1.5$, avec une marge d'erreur de 0.05 g/L.\\
Une concentration autour de 0.3 g/L dans les gouttes de bactéries
correspond à une concentration molaire de 0.3 mM, largement en dessous
de la concentration molaire de 5 M nécessaire pour induire un effet de
gonflement comparable avec des gouttes de glucose (50\,\%) d\'eposées sur
le même gel (la molécule de glucose est environ cinq fois plus petite
que la molécule de surfactine). Un simple gradient de concentration ---
entre la goutte contenant de la surfactine et le gel agar n'en
contenant pas --- avec l'interface gel--eau agissant comme une membrane
semi-perméable ne semble pas suffire pour expliquer le mécanisme de
pompage car dans ce cas la théorie prédit que la pression osmotique ne
dépend que marginalement de l'espèce chimique et principalement de sa
concentration molaire. Il est possible que le pompage fasse intervenir
une interaction spécifique entre les molécules de surfactine et la
matrice de gel d'agar, capable de libérer le solvant et d'induire un
accroissement du volume de gouttes deposées en surface. Une
observation qui va dans le sens d'une telle interaction est que quand
la surfactine ou les bactéries en produisant sont ajoutées directement
au gel, celui-ci, soit ne se solidifie plus, soit se liquéfie après
quelque temps.\\

\begin{figure*}[ht!]
\centering
\includegraphics[width=0.9\textwidth]{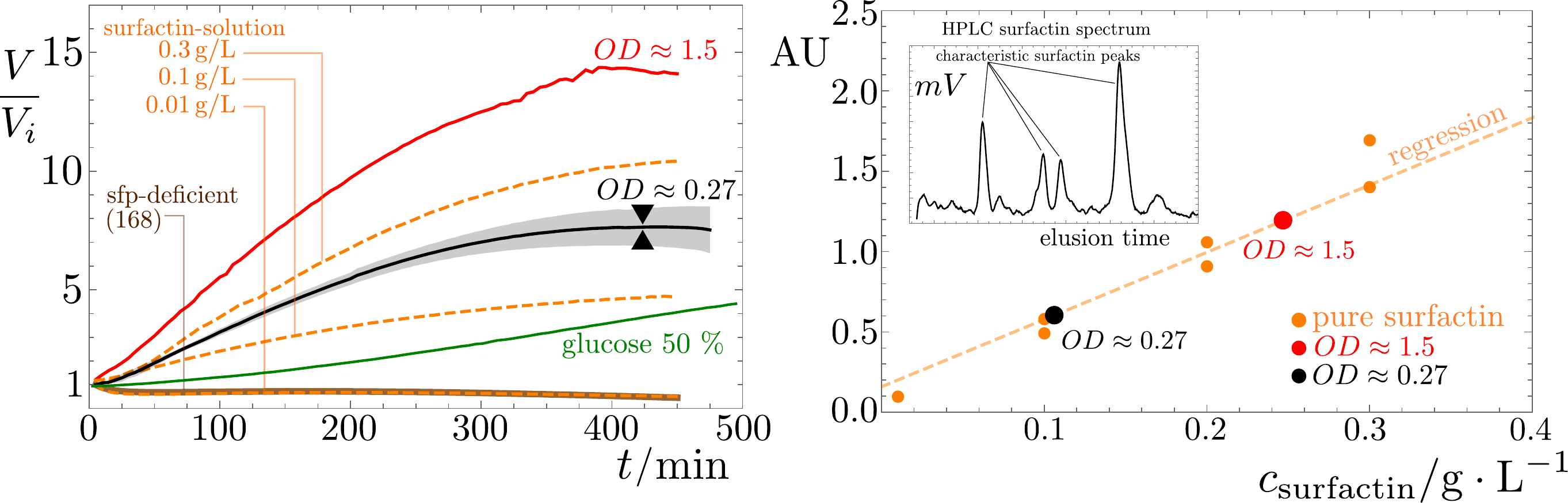}
\caption{(a) Des gouttes de surfactine pure
  d'une concentration entre 0.1 et 0.5 g/L (0.1--0.5 mM pour une masse molaire de 1030 g/mol) affichent un gonflement
  similaire aux gouttes de bactéries d'une OD de 0.27. Une augmentation de volume est observable pour des gouttes de glucose (50\,\%) à une concentration molaires dix mille fois plus grande (5 M). (b)
  La concentration en surfactine des gouttes de bactéries est estimée
  à 0.1--0.3 g/L par spectrométrie HPLC. La concentration est obtenue en
  calculant la superficie en unités arbitraires (AU) des pics
  caractéristiques d'absorbtion de la surfactine dans le spectre.}
\end{figure*}

\begin{figure*}[ht!]
\centering
\includegraphics[width=0.5\textwidth]{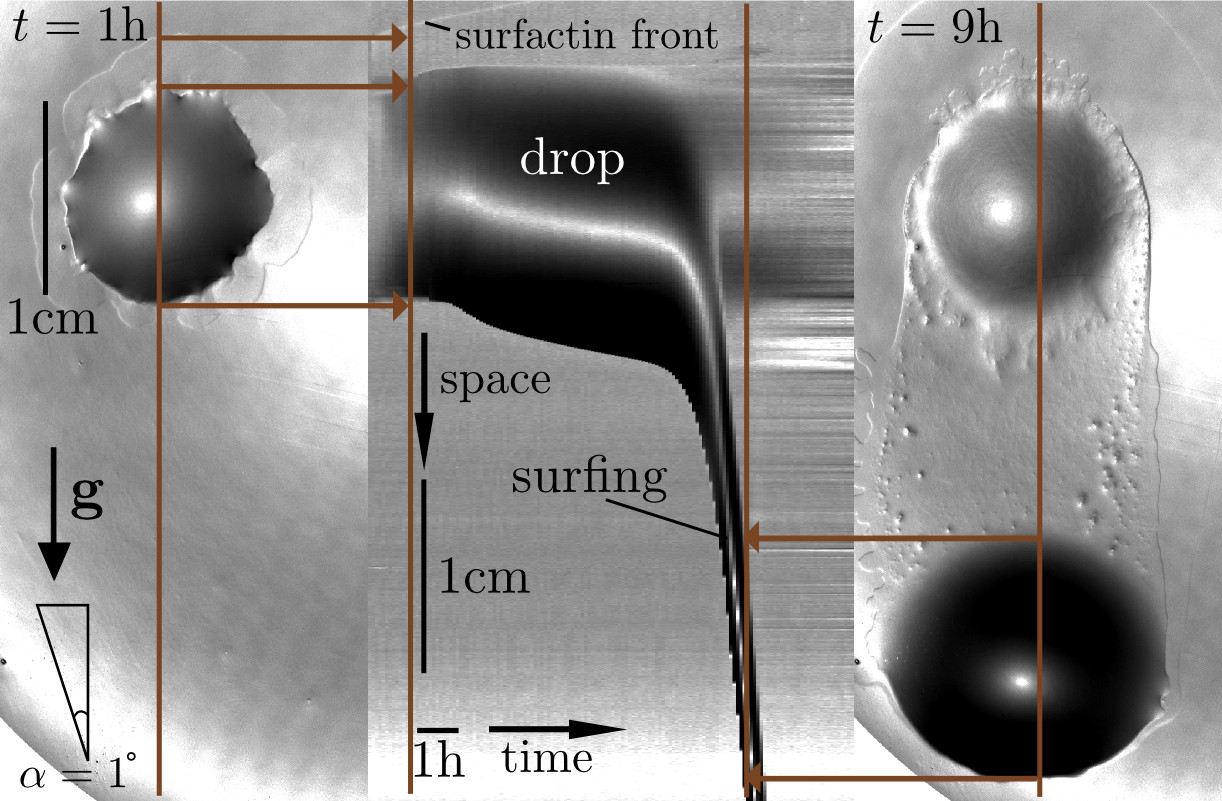}
\caption{Pour une goutte de bactéries ($\mathrm{OD} = 0.27, V_i = 2$ $\mu$L) déposée sur un gel d'agar légèrement incliné ($1^{\circ}$), l'accroissement du volume en combinaison avec les propriétés tensio-actifs de la surfactine permet à celle-ci de dévaler le plan 7 à 8 heures après d\'ep\^ot.}
\end{figure*}

\noindent
\textbf{Discussion}\\
Dans cet article, nous avons etudié un nouveau mécanisme
physico-chimique permettant à des gouttes de bactéries en milieu
partiellement humide de croître en volume. Parmi les divers molécules
produites et exportées par les bactéries, nous avons identifié la
surfactine, un tensio-actif puissant, comme étant responsable de cet
effet. Nous pensons que ce mécanisme peut avoir un rôle immense pour
{\em B. subtilis} dans son environnement naturel comme les sols
non saturés  \cite{dani2007}, d'une part en lui permettant en cas de
sécheresse d'extraire de l'eau de l'environnement, et d'autre part en
lui permettant de mieux migrer. Vitesses et distances de migration
sont susceptibles d'augmenter avec la quantité d'eau disponible, que
ce soit dans les milieux poreux  \cite{sadjadi2013} ou à la surface d'un
substrat comme l'agar  \cite{leclere2006,julkowska2005,henrichsen1972}.
Pour les gouttelettes étudiées ici, le volume gagné augmente la
traction gravitationelle par rapport au forces capillaires et permet
de franchir le seuil d'étalement ou de dévalement sur des substrats
non horizontaux. Sur des gels d'agar, les gouttes de bactéries
glissent ainsi sur des pentes de l'ordre du degré (cf. Fig. 4), quand des gouttes d'eau de même volume initial restent accrochées même à la verticale, piégées par les forces capillaires.

\end{document}